\pdfoutput=1
\documentclass[11pt]{article}

\usepackage{acl}

\usepackage{times}
\usepackage{latexsym}

\usepackage[T1]{fontenc}
\usepackage[utf8]{inputenc}

\usepackage{microtype}

\usepackage{inconsolata}

\usepackage{graphicx}

\title{GNN-Coder: Boosting Semantic Code Retrieval with Combined GNN and Transformer}

\author{
 \textbf{Yufan Ye\textsuperscript{1}},
 \textbf{Pu Pang\textsuperscript{2}},
 \textbf{Ting Zhang\textsuperscript{3}},
 \textbf{Hua Huang\textsuperscript{3,}\thanks{Corresponding author}}
\\
\\
 \textsuperscript{1}Beijing Institute of Technology,
 \textsuperscript{2}Xi’an Jiaotong University,
 \textsuperscript{3}Beijing Normal University
\\
  \href{}{yeyufan@bit.edu.cn}, \href{}{fankewen@stu.xjtu.edu.cn}, \href{}{
  \{tingzhang, huahuang\}@bnu.edu.cn}
}

\usepackage{amsmath}
\usepackage{adjustbox}
\usepackage{arydshln}
\usepackage{caption}
\usepackage{subcaption}
\usepackage{multirow}

\begin{document}
\maketitle
\begin{abstract}
Code retrieval is a crucial component in modern software development, particularly in large-scale projects.
However, existing approaches relying on sequence-based models often fail to fully exploit the structural dependencies inherent in code, leading to suboptimal retrieval performance, particularly with structurally complex code fragments. In this paper, we introduce GNN-Coder, a novel framework based on Graph Neural Network (GNN) to utilize Abstract Syntax Tree (AST). We make the first attempt to study how GNN-integrated Transformer can promote the development of semantic retrieval tasks by capturing the structural and semantic features of code.
We further propose an innovative graph pooling method tailored for AST, utilizing the number of child nodes as a key feature to highlight the intrinsic topological relationships within the AST. This design effectively integrates both sequential and hierarchical representations, enhancing the model's ability to capture code structure and semantics.
Additionally, we introduce the Mean Angular Margin (MAM), a novel metric for quantifying the uniformity of code embedding distributions, providing a standardized measure of feature separability. The proposed method achieves a lower MAM, indicating a more discriminative feature representation. This underscores GNN-Coder's superior ability to distinguish between code snippets, thereby enhancing retrieval accuracy.
Experimental results show that GNN-Coder significantly boosts retrieval performance, with a 1\%-10\% improvement in MRR on the CSN dataset, and a notable 20\% gain in zero-shot performance on the CosQA dataset.

 % By utilizing a small retrieval model to provide context information, code retrieval can help software developers efficiently reuse existing code resources. 
% At the same time, the bias generated in the pre-trained model due to the use of large-scale and diverse training data has an impact on the retrieval performance on small code libraries. To address the above problems, we propose GNN-Coder, a novel framework based on Graph Neural Networks (GNNs), which strengthens the code embedding of AST through a hierarchical structure. On this basis, we further design a new graph pooling method. This method is specifically tailored for AST and calculates the importance score by using the number of child nodes rather than their characteristics. This small hierarchical GNN model can handle all nodes in the AST. By integrating more AST information and reducing bias, it enhances the embedding from the Transformer model. In addition, to evaluate the bias, we propose a new metric, the Mean Allocation Score (MMS). Experimental results show that GNN-Coder effectively reduces the bias. The MMR on the CSN dataset is increased by 1\%-9.5\%, and the improvement range on the CosQA dataset reaches 20\%-22\%. 
\end{abstract}

\section{Introduction}
Code retrieval, as a technology that takes natural language queries as input and outputs code snippets that match the query intent, plays a facilitating role in program reuse during the software development process~\citep{li2022coderetriever,liu2024empirical}. Meanwhile, it also drives the latest research in the field of retrieval-augmented generation~\citep{zhou2022docprompting,wang2024searching}. The main challenge faced by effective code retrieval lies in the semantic gap between natural language descriptions and source code. This is because natural language descriptions and source code are heterogeneous resources that share very few lexical tokens, synonyms, and language structures~\citep{gu2018deep,zhu2022survey}.

With the Transformer achieving success in the field of NLP\citep{achiam2023gpt,grattafiori2024llama}, CodeBERT treats code as a special form of natural language. Specifically, it processes code snippets and their corresponding texts as pairs of natural language sentences in the BERT architecture~\citep{devlin2018bert}. However, code has unique syntax and structure different from natural languages, such as AST~\citep{zhang2019novel,tang2021ast}, which reflects the hierarchical organization of code. 
% Although the attention mechanism of the Transformer can calculate the relationships between all tokens~\citep{vaswani2017attention}, it ignores the inherent structured sparse relationships in AST. 
To improve the performance of the model, researchers have made efforts~\cite{guo2020graphcodebert,wang2021codet5,wang2023codet5+} to integrate AST into the transformer training process. 
For example, UniXcoder~\citep{guo2022unixcoder} introduces a one-to-one mapping function to convert the AST into a sequence. 
However, these methods directly flatten the components of AST to be a linear sequence, ignoring its inherent structural potential and impairing performance in tasks that require deep syntactic understanding.

% Some methods manually extract grammatical details by leveraging specific components of the AST. For example, GraphCodeBERT~\citep{guo2020graphcodebert} utilizes the data flow derived from the AST, while CodeT5~\citep{wang2021codet5} and CodeT5+~\citep{wang2023codet5+} focus on the identifier nodes in the AST. There are also some methods that attempt to use the complete AST. For instance, UniXcoder~\citep{guo2022unixcoder} introduces a one-to-one mapping function to convert the AST into a sequence. However, these methods have certain limitations in making full use of grammatical information in the AST. The reason is either that the loss function cannot directly supervise the complete AST, or that the AST is too long to be input as a sequence.

\begin{figure}[t]
  \centering
  \includegraphics[width=0.5\textwidth]{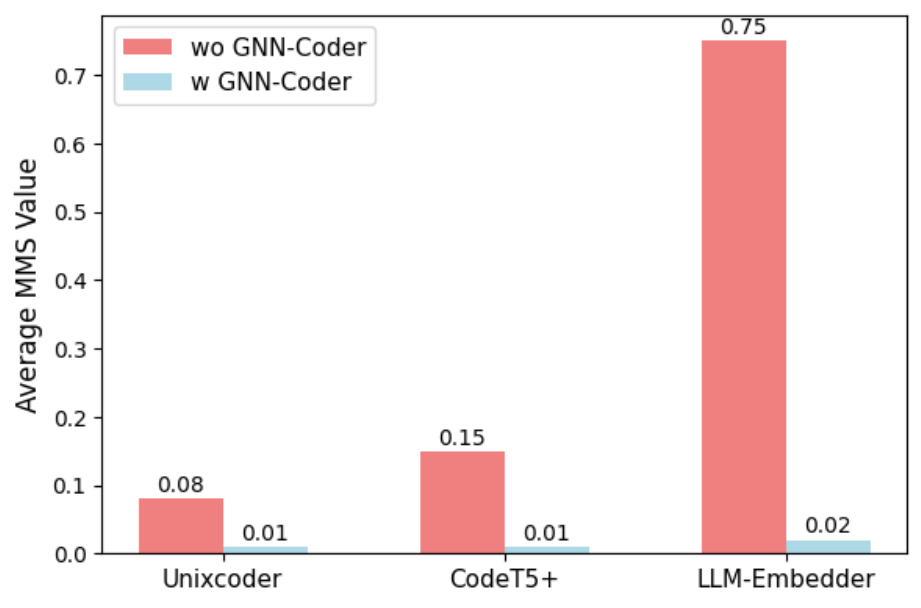}
  \caption{The average MAM for six PLs in CSN dataset. A value close to 0 indicates thorough feature separation.}
  \label{fig:mam}
\end{figure}

In this paper, we propose a novel GNN-based framework, investigating the first attempt to explore how GNN-integrated Transformer make full use of the complete syntactic information of the AST for the code retrieval task.
% , and at the same time effectively alleviate the biases existing in the Transformer model. 
The inherent sparsity of GNNs makes them particularly suitable for handling tree-structured AST. 
% Moreover, the GNN model can also serve as a mapping function to transform the biased code embeddings into an unbiased space. 
Specifically, we begin by encoding the content of AST nodes using a frozen Transformer model, while representing their types through one-hot encoding. Next, we employ a GNN model to integrate information across all AST nodes and generate a unified code embedding. To align the code and text embeddings, we apply contrastive loss during the training of the GNN model. Considering the tree-structured nature of the AST and the encoding requirements, we propose a hierarchical GNN model incorporating a graph pooling layer. Additionally, we introduce a novel graph pooling method, ASTGPool, designed specifically for AST, which emphasizes intrinsic topological relationships within the AST and accelerates information propagation to the root node, thereby improving retrieval accuracy.

To further evaluate the discriminative power of learned code features, we introduce the Mean Angular Margin (MAM) metric. MAM calculates the cosine similarity between a given text embedding and all code embeddings, offering a robust measure of feature differentiation—an aspect often neglected in prior research, yet essential for effective code retrieval. The ideal scenario, as indicated by MAM, occurs when the cosine similarities between distinct code embeddings approach zero, reflecting thorough separation of code embeddings. 
% This separation is vital for ensuring that code representations are sufficiently discriminative, capturing subtle differences in code structure and semantics. 
As shown in Figure~\ref{fig:mam}, GNN-Coder achieves a lower MAM value, demonstrating improved feature separation.
Experimental results show that GNN-Coder significantly boosts retrieval performance, with a 1\%-10\% improvement in MRR on the CSN dataset, and a notable 20\% gain in zero-shot performance on the CosQA dataset.

% In addition, although large pre-trained datasets are helpful for improving model performance, in retrieval tasks, especially in the scenario of small code databases, they may introduce biases.~\citep{zhang2023code} To fully capture the semantics of code, pre-trained datasets are usually large in scale and cover a variety of programming languages (PLs). This characteristic makes it easy for pre-trained models to produce biases when encoding code snippet databases for a single PL. Since it is quite difficult to directly measure the distribution of code embeddings, and traditional methods often overlook the characteristics of retrieval tasks, we propose a new method to evaluate whether their distribution is uniform. Assuming that the code embeddings have reached a uniform distribution, the average value of the cosine similarity between any given text embedding and all code embeddings should theoretically be zero. We define this metric as the Mean Matching Score (MMS). By analyzing the mean value of MMS for all texts, the uniformity of code embeddings can be evaluated. As shown in Figure~\ref{fig:mms}, when encoding the part of a single PL in the CSN dataset, UniXcoder and CodeT5+ exhibit slight biases, while LLM-Embedder~\citep{zhang2023retrieve}, a general NLP retrieval model, shows more obvious biases. These biases hinder further improvement of the model's retrieval performance.

In summary, our contributions are as follows:
\begin{itemize}
\item We propose a novel framework based on GNN for code retrieval, leveraging the structured sparse relationship captured by AST. To the best of our knowledge, this is the first attempt to explore the AST-guided GNN in Transformer module for code retrieval task.

% This framework is built based on GNNs. By integrating AST information, it effectively enhances the code embedding encoded by the Transformer model. Experimental results show that even if the Transformer model has been pre-trained using part of the AST, GNN-Coder can still significantly improve the performance of the model in code retrieval task.

\item We introduce a novel graph pooling method, ASTGPool, tailed for AST. The design evaluates the importance score by utilizing the number of child nodes, which characterize the topological relationship. We show that the proposed pooling method is superior to the existing pooling techniques.

\item Experiments validate the effectiveness of the proposed framework. The introduced GNN module enhances the performance of the model which has leveraged part of the AST as a linear sequence. Additionally, we present a new metric, MAM, to assess code representation distribution, demonstrating that the learned features are effectively separated, benefiting code retrieval. 
% propose a novel interactive metric method called MMS, which is used to evaluate the uniformity of the embeddings. The research finds that the Transformer model has a bias when encoding specific code databases, while GNN-Coder can effectively alleviate this bias.
\end{itemize}

\begin{figure*}[t]
  \centering
  \includegraphics[width=1.0\textwidth]{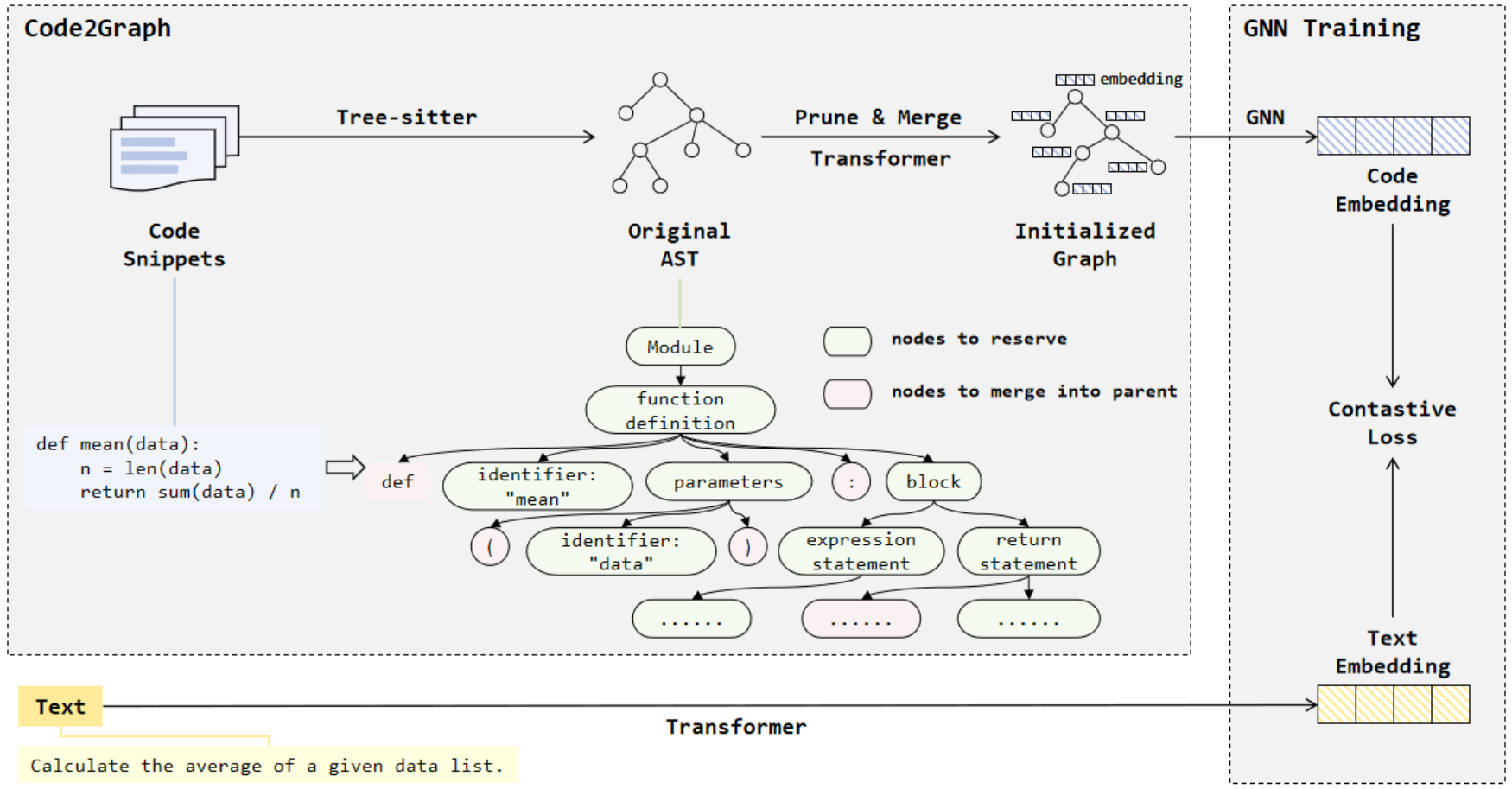}
  \caption{Overall architecture of GNN-Coder. The code is transformed into an AST, which is initialized with a Transformer model, processed by a GNN, and aligned with text embeddings through a contrastive loss function.
}
  \label{fig:overall_architecture}
\end{figure*}

\section{Related Work}
\noindent 
\textbf{Code Retrieval.}
% \subsection{Code Retrieval}
Software developers often rely on the reuse of existing code resources to achieve efficient code. Therefore, code retrieval has emerged as a crucial research area, aiming to explore the implicit connections between natural language queries and code databases, enabling developers to obtain the required code quickly and accurately. Early research mainly represent code and queries as feature vectors and retrieve code based on similarity to the query vectors. Boolean vectors~\citep{salton1983extended} characterize the features by indicating the presence or absence of specific features, such as specific types of AST nodes~\citep{luan2019aroma}. Another commonly used method is to map a set of tokens into a Term Frequency-Inverse Document Frequency (TF-IDF) vector, which can not only indicate whether a feature exists but also reflect the importance of this feature~\citep{diamantopoulos2018codecatch,takuya2011spontaneous}.  

With the rapid development of deep learning technology, an increasing number of studies have focused on the use of neural networks to achieve efficient code retrieval. Most of the related work adopts end-to-end neural learning methods. The query and the code are embedded into a joint vector space through the model, and the code search problem is then transformed into finding the nearest-neighbor code for a given query in this space~\citep{gu2018deep,sun2022code}. 
The core of code retrieval lies in code encoding, which is reviewed below.
% Subsequent research has been further dedicated to improving the vector representation of code learning. 

\noindent 
\textbf{Code Encoding With Transformer.}
% \subsection{Code Encoding With Transformer}
Following the significant success of the transformer architecture and pre-training in NLP, researchers have started to explore the potential of the Transformer model in code representation learning. Single encoder models mainly follow the BERT framework. CodeBERT~\citep{feng2020codebert} utilizes the masked language objective to pre-train on NL-PL pairs, and adds the substitute token detection task~\citep{clark2020electra}. GraphCodeBERT~\citep{guo2020graphcodebert} enhances the pre-training process by incorporating the data flow derived from the AST. SynCoBERT~\citep{wang2021syncobert} utilizes identifier prediction, AST edge prediction, and multimodal contrastive learning to make full use of AST. In addition, some works adopt the encoder-decoder architecture. 
% PLBART~\citep{ahmad2021unified} adjusts the BART~\citep{lewis2019bart} architecture and uses the denoising objective to pre-train on NL and PL corpora. 
CodeT5~\citep{wang2021codet5} extends the T5~\citep{raffel2020exploring} model to the code domain and additionally combines the information of the identifier nodes in the AST. TreeBERT~\citep{jiang2021treebert} utilizes the structural information of the AST by representing it as a set of paths from the root node to the terminal nodes. 

The aforementioned methods utilize Transformer models for code encoding. While some integrate AST information, they still treat it as a linear sequence. This sequence-based representation hampers the accurate capture of hierarchical and complex relationships inherent in code structures. Consequently, these approaches fail to fully leverage the syntactic and semantic richness of programming languages, which are defined by intricate nested structures that cannot be effectively represented in a flat sequential format.

% When perceiving the topological information of complex graph structures like AST, the Transformer performs relatively weakly. Although recent studies have employed means such as additional positional encoding to represent structural information, it still cannot capture the local and global structural dependency relationships between nodes as accurately as GNN. 

\noindent 
\textbf{Code Encoding With GNN.}
% \subsection{Code Encoding With GNN}
\citet{allamanis2017learning} first use GNN to represent code. To convert the code into a graph, they design complex edges based on the AST. Devign~\citep{zhou2019devign} and Reveal~\citep{chakraborty2021deep} adopt the Code Property Graph (CPG), which is composed of the AST, the control flow graph (CFG) and the data flow graph (DFG), and used the Gated Graph Neural Network (GGNN) to encode the graph. 
% IVDetect~\citep{li2021vulnerability} further enriched the content of nodes by adding more information, such as AST subtrees, data dependency contexts, and control dependency contexts.
Graphsearchnet~\citep{liu2023graphsearchnet} introduces Bidirectional GGNN (BiGGNN) to create graphs for code and text, capture local structural details, and uses a multi-head attention module to enhance BiGGNN.

These methods mainly use GNN as a tool for encoding code for vulnerability identification. In our research work, we focus on the multimodal retrieval task and integrate Transformer with GNN.
% use contrastive learning to align the embeddings of code and text. 
To ensure that our model is applicable to various PLs, we use only the basic AST without adding complex edges or nodes. 

% The above-mentioned methods either use Transformer or GNN alone, failing to fully integrate the advantages of both. In our research, we combine the two and propose an innovative framework, focusing on improving semantic code retrieval ability. 

\section{Methodology}
\subsection{Overall Architecture}
The architecture of GNN-Coder is depicted in Figure~\ref{fig:overall_architecture}. In the Code2Graph stage, we utilize tools to convert the code into an AST, which is then initialized using a Transformer model. The initialized graph is processed by a GNN model. Finally, a contrastive loss is applied to align the embedding representations of both code and text.

\subsection{Code2Graph}
\textbf{Code2AST.} We adopt the same method as in the code Transformer series of works and use Tree-sitter\footnote{https://github.com/tree-sitter/tree-sitter} to convert various PLs into ASTs. 
% An example of an AST generated by Tree-sitter is shown in Figure~\ref{fig:overall_architecture}.

\noindent
\textbf{Refine AST.} 
%Compared with the ASTs generated by Joern\footnote{https://github.com/joernio/joern}, which is commonly used in graph-based models, 
The ASTs generated by Tree-sitter are usually more redundant. Therefore, we need to further simplify the AST. Specifically, we delete the "shadow nodes" with the same type and content. However, directly deleting these nodes may lead to the loss of code syntax information. To solve this problem, we merge these nodes into their parent nodes. The specific operation is as follows: we reconstruct the content of the parent node by concatenating the contents of all child nodes, except for the "block" node. The "block" node, as a container, has its content sourced from the contents of all child nodes. For example, the "function\_definition" node in Figure~\ref{fig:overall_architecture} needs to be reconstructed because it has child nodes to be deleted (i.e., "def" and ":"). First, we replace the content of the "function\_definition" node with the concatenation of the contents of all child nodes except the "block" node. Then, we delete the "def" and ":" nodes. Eventually, the content of the "function\_definition" node becomes "def mean (data):", which includes the information of the deleted nodes.

\noindent
\textbf{AST2Graph.} To meet the input requirements of GNN, we need to represent the types and contents of AST nodes as vectors. Previous methods~\citep{devlin2018bert,chakraborty2021deep,li2021vulnerability} used one-hot encoding and a pre-trained word2vec model to encode the types and contents of AST nodes respectively, and then concatenated these encodings to form the initial node embeddings. In this research work, we replace the word2vec model with a pre-trained Transformer model to fully utilize the powerful ability of the Transformer model in encoding context information. In addition, referring to methods such as Devign, we reverse the direction of the edges in the original AST to ensure that the information of the leaf nodes can be effectively propagated to the root node. 

\begin{figure}[t]
  \centering
  \includegraphics[width=0.5\textwidth]{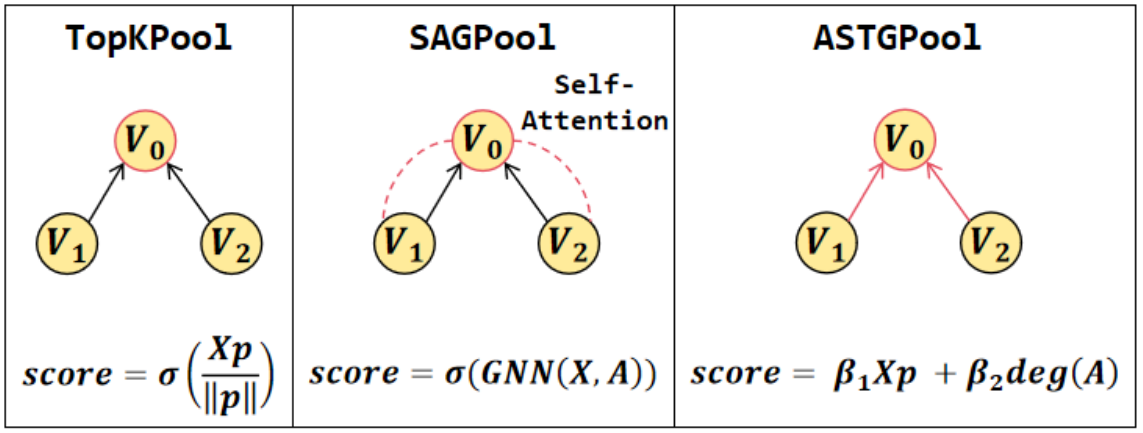}
  \caption{Illustrating importance score calculation for different pooling methods. ``deg'' represents in-degree.}
  \label{fig:pooling}
\end{figure}

\begin{figure*}[t]
  \centering
  \includegraphics[width=1.0\textwidth]{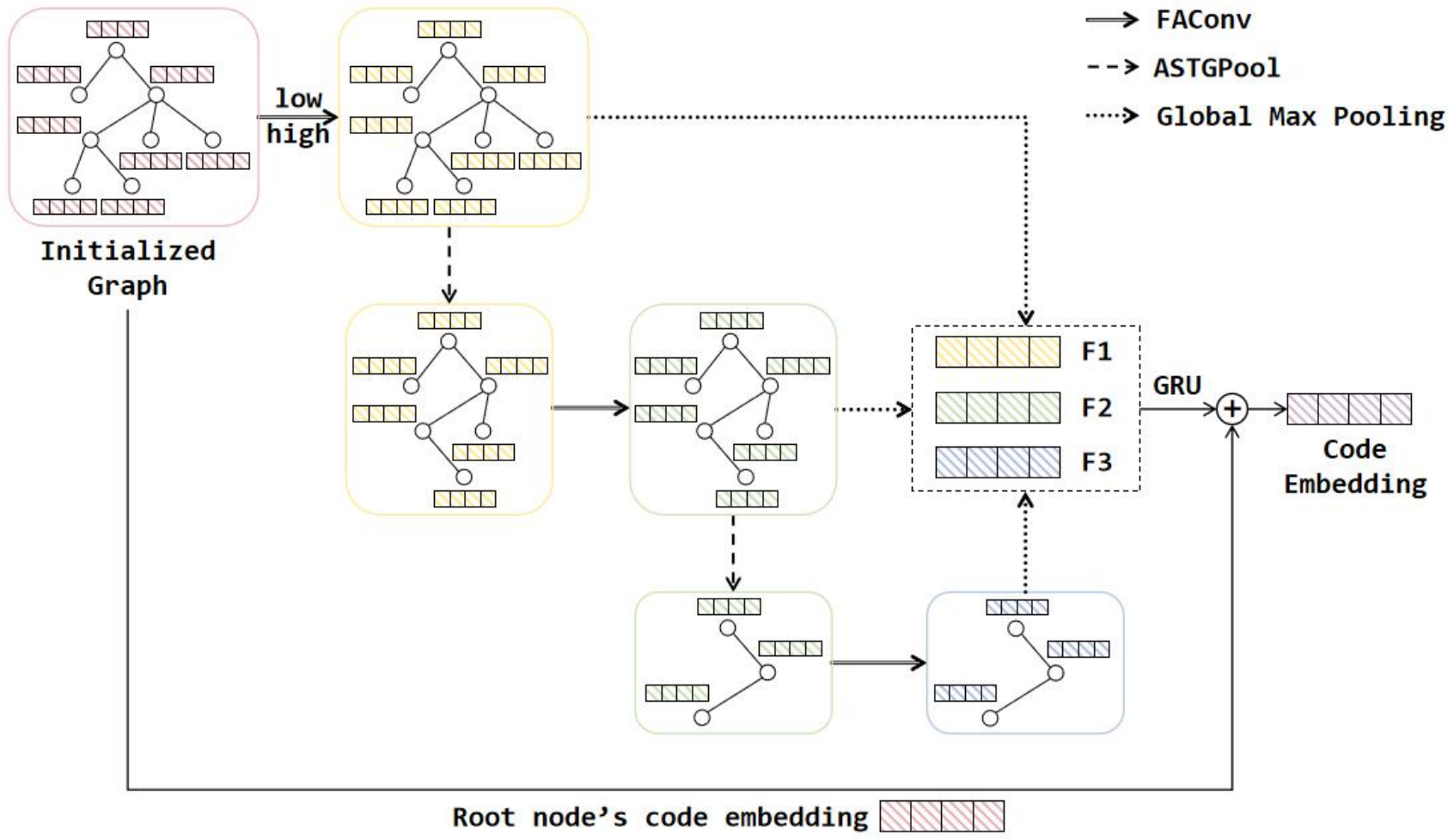}
  \caption{Illustrating the GNN model, which is a hierarchical architecture incorporated with ASTGPool layer. Here we show a hierarchical depth of 3 and $F1,F2,F3$ represent the features extracted at each corresponding depth.}
  \label{fig:gnn_model}
\end{figure*}

\subsection{GNN Training}

\subsubsection{GNN Architecture}
\label{sec:GNN_architecture}
Since the root node in the tree structure contains much richer information, conventional GNN models struggle to handle it effectively. In view of this, we meticulously design a hierarchical GNN model that incorporates a novel graph pooling layer, adapting to the AST and its coding requirements.

\noindent
\textbf{Conv Layer.} 
The GNN model we construct employs the FAConv layer~\citep{bo2021beyond} as the convolutional layer. The FAConv layer can adaptively adjust the coefficients of low-frequency and high-frequency signals without prior knowledge of the network type. This remarkable flexibility makes it suitable for processing graphs initialized in different ways. 

\noindent
\textbf{Graph Pooling.} 
We propose a hierarchical architecture that incorporates a graph pooling layer into the original FAGCN model. This design enables the capture of information at multiple granularity levels. The graph pooling layer plays a key role by accelerating information aggregation from leaf nodes to the root node and effectively filtering noise in the leaf nodes.
% , while significantly reducing the GPU memory consumption. 

To explore the most suitable graph pooling method for AST, 
we experimentally study TopKPool~\citep{gao2019graph,cangea2018towards} and SAGPool~\citep{lee2019self,knyazev2019understanding}. However, our findings indicate that these methods are not fully appropriate for AST. Specifically, while SAGPool computes importance scores based on the features of neighboring nodes, its performance is sometimes inferior to TopKPool, which relies solely on node features. This discrepancy arises because the inversion of edges repositions child nodes as neighbors, introducing noise into the parent node's importance calculation. Based on the above situation, we propose a novel graph pooling method specifically designed for AST, called ASTGPool. It evaluates the importance of nodes based on node features and the number of adjacent nodes, which emphasizes intrinsic topological relationships. 
The formula for calculating the importance score is as follows:
\begin{equation}
score = \beta_1 X p + \beta_2 deg(A),
\end{equation}
where $X$ denotes the node feature matrix, $A$ is the adjacency matrix, and $\beta_1$, $\beta_2$ are learnable parameters that balance the contribution of each factor. Figure~\ref{fig:pooling} illustrates the importance score calculation for different pooling methods.

\noindent
\textbf{Overall Architecture.} 
The overall architecture consists of stacking the FAConv and ASTGPool layers $L$ times, with a global maximum pooling layer added after each FAConv layer to capture embeddings at different granularities.
To effectively fuse these multi-level embeddings, we concatenate them and then generate the final code embeddings through a Gated Recurrent Unit (GRU)~\citep{cho2014learning}. 
The root node contains the source code in text format. To preserve the Transformer model's context encoding capabilities and expedite training, we incorporate a residual connection between the root node's code embeddings and the final code embeddings, following the design principles of the CLIP-Adapter~\citep{gao2024clip}. The GNN model architecture is detailed in Figure~\ref{fig:gnn_model}.

\subsubsection{Training objective} 
We draw on the contrastive representation learning method employed in CLIP~\citep{radford2021learning} to align the code embeddings generated by the GNN model with the text embeddings generated by the Transformer model. This method is concise and efficient, enabling the learning of aligned embeddings across different modalities. The formula for the loss function used to align the code embeddings and the text embeddings is as follows:
\begin{align}
\mathcal{L}_{\text{Code}} &= - \frac{1}{N} \sum_{i=1}^N \log \frac{\exp \left( \text{sim}(\mathbf{c}_i, \mathbf{t}_i) / \tau \right)}{\sum_{j=1}^N \exp \left( \text{sim}(\mathbf{c}_i, \mathbf{t}_j) / \tau \right)}, \nonumber \\
\mathcal{L}_{\text{Text}} & = - \frac{1}{N} \sum_{i=1}^N \log \frac{\exp \left( \text{sim}(\mathbf{c}_i, \mathbf{t}_i) / \tau \right)}{\sum_{j=1}^N \exp \left( \text{sim}(\mathbf{c}_j, \mathbf{t}_i) / \tau \right)}, \nonumber\\
\mathcal{L} &= \frac{1}{2} \left( \mathcal{L}_{\text{Code}} + \mathcal{L}_{\text{Text}} \right).
\end{align}
Here, $\mathbf{c}_i$ and $\mathbf{t}_i$ represent the $i$-th pair of code embedding and text embedding, respectively. The function $\text{sim}(\cdot, \cdot)$ represents the cosine similarity between two vectors. The parameter $\tau$ is a learnable temperature parameter.

\section{Experiments}

\begin{table*}[t]
\centering
\renewcommand{\arraystretch}{1.0}
\begin{adjustbox}{width=\textwidth}
\begin{tabular}{l|cc|cc|cc|cc|cc|cc|c}
\hline

Model & \multicolumn{2}{c|}{Ruby} & \multicolumn{2}{c|}{JavaScript} & \multicolumn{2}{c|}{Go} & \multicolumn{2}{c|}{Python} & \multicolumn{2}{c|}{Java} & \multicolumn{2}{c|}{PHP} & MRR\_Avg \\
 & MRR & R@1 & MRR & R@1 & MRR & R@1 & MRR & R@1 & MRR & R@1 & MRR & R@1 \\
\hline
UniXcoder 110M & 45.02 & 34.26 & 34.28 & 25.34 & 54.64 & 43.61 & 33.31 & 24.42 & 36.28 & 26.71 & 24.94 & 17.51 & 38.08  \\
\ + GNN-Coder & \textbf{50.40} & \textbf{40.05} & \textbf{40.22} & \textbf{30.42} & \textbf{67.76} & \textbf{58.04} & \textbf{44.56} & \textbf{34.37} & \textbf{46.42} & \textbf{35.60} & \textbf{35.86} & \textbf{26.46} & \textbf{47.54}  \\
\hdashline
CodeT5+ 110M & 73.55 & 64.55 & 65.83 & 56.24 & 89.51 & 84.55 & 69.75 & 60.12 & 69.42 & 59.74 & 64.44 & 54.06 & 72.08 \\
\ + GNN-Coder & \textbf{73.85} & \textbf{64.71} & \textbf{67.20} & \textbf{57.70} & \textbf{90.71} & \textbf{86.17} & \textbf{70.37} & \textbf{60.77} & \textbf{70.80} & \textbf{61.46} & \textbf{65.93} & \textbf{55.47} & \textbf{73.14} \\
\hline
LLM-Embedder & 63.07 & 53.29 & 49.33 & 39.65 & 80.94 & 73.02 & 55.86 & 45.52 & 53.60 & 42.94 & 44.70 & 34.56 & 57.92  \\
\ + GNN-Coder & \textbf{65.00} & \textbf{55.19} & \textbf{52.35} & \textbf{42.27} & \textbf{87.10} & \textbf{81.30} & \textbf{61.71} & \textbf{51.47} & \textbf{60.38} & \textbf{49.72} & \textbf{54.96} & \textbf{44.19} & \textbf{63.58}  \\
\hline

\end{tabular}
\end{adjustbox}
\caption{Results comparison in terms of MRR and R@1 on various Transformer architectures on the CSN dataset, showing GNN-Coder consistently enhances the performance of all Transformer-based models.}
\label{tab:csn_mrr_recall}
\end{table*}

\subsection{Settings}
\label{sec:dataset_and_metric}

\noindent 
\textbf{Datasets.}
In code retrieval task, we use the CSN and CosQA datasets to evaluate the performance of GNN-Coder. The CSN is derived from the CodeSearchNet dataset~\citep{husain2019codesearchnet}, and it filters out low-quality queries through manually crafted rules~\citep{guo2020graphcodebert}. CosQA~\citep{huang2021cosqa}, on the other hand, uses queries from the logs of the Microsoft Bing search engine and the corresponding code snippets. The experimental results of the CSN dataset can demonstrate the basic performance of GNN-Coder in code retrieval, while the results of the CosQA dataset can verify its generalization ability, as the queries in the CosQA dataset are not included in the training set. 

\noindent
\textbf{Baselines.}
Our GNN architecture is applicable to various Transformer models.
It is crucial to evaluate its compatibility and adaptability with various state-of-the-art Transformer models.
Hence, to comprehensively evaluate the performance of GNN-Coder, we select a variety of Transformer models, including UniXcoder, CodeT5+, and LLM-Embedder. UniXcoder and CodeT5+ represent state-of-the-art advancements in code retrieval, while LLM-Embedder is tailored to address the unique retrieval enhancement needs of LLMs.
% By employing these different types of Transformer models, we aim to explore the adaptability of GNN-Coder in a diverse model environment.

\noindent
\textbf{Evaluation Metrics.}
We adopt Mean Reciprocal Rank (MRR) and Recall@K~\citep{liu2021opportunities,di2023code} as the retrieval metrics. 
In addition, we utilize the proposed MAM to evaluate the distribution of code embeddings. MAM are calculated through the average cosine similarity between text embeddings and all code embeddings. The specific formula is as follows:
\begin{equation}
\text{MAM}_j = \frac{1}{N} \sum_{i=1}^{N} \text{sim}(\mathbf{c}_i, \mathbf{t}_j).
\end{equation}
By analyzing the distribution of all MAM, we can evaluate the uniformity of the code embedding distributions. Theoretically, if the code embeddings are uniformly distributed, both their mean and standard deviation (SD) should be close to zero. 
% A non-zero mean indicates a bias in the distribution of code embeddings. 

\subsection{Implementation Details}
For UniXcoder and CodeT5+, we use the configurations provided by the authors to encode code and text. For LLM-Embedder, without additional instructions, we use the embedding of the first token in the last hidden layer as the encoded embeddings.

Regarding the configuration of the GNN model, we set it to three layers. The size of the hidden layer of the GNN model is dynamically adjusted according to the output dimension of the Transformer model and the number of AST node types. Meanwhile, the output dimension of the GNN model is kept consistent with that of the corresponding Transformer model. More details related to the training process can be found in Appendix~\ref{app:training_process}. 

\subsection{Results}

\begin{table*}[t]
\centering
\renewcommand{\arraystretch}{1.0}
\begin{adjustbox}{width=\textwidth}
\begin{tabular}{l|cc|cc|cc|cc|cc|cc}
\hline

Model & \multicolumn{2}{c|}{Ruby} & \multicolumn{2}{c|}{JavaScript} & \multicolumn{2}{c|}{Go} & \multicolumn{2}{c|}{Python} & \multicolumn{2}{c|}{Java} & \multicolumn{2}{c}{PHP} \\
 & Mean & SD & Mean & SD & Mean & SD & Mean & SD & Mean & SD & Mean & SD \\
\hline
UniXcoder 110M & 0.09 & 0.02 & 0.09 & 0.02 & 0.07 & 0.02 & 0.08 & 0.02 & 0.08 & 0.02 & 0.08 & 0.02 \\
\ + GNN-Coder & \textbf{0.01} & \textbf{0.01} & \textbf{0.01} & 0.02 & \textbf{0.01} & \textbf{0.01} & \textbf{0.00} & \textbf{0.01} & \textbf{0.01} & \textbf{0.01} & \textbf{0.00} & 0.02 \\
\hdashline
CodeT5+ 110M & 0.16 & 0.05 & 0.17 & 0.05 & 0.15 & 0.03 & 0.13 & 0.03 & 0.14 & 0.04 & 0.15 & 0.04 \\
\ + GNN-Coder & \textbf{0.01} & \textbf{0.01} & \textbf{0.02} & \textbf{0.01} & \textbf{0.00} & \textbf{0.01} & \textbf{0.01} & \textbf{0.01} & \textbf{0.01} & \textbf{0.01} & \textbf{0.01} & \textbf{0.01} \\
\hline
LLM-Embedder & 0.76 & 0.01 & 0.77 & 0.01 & 0.69 & 0.02 & 0.76 & 0.01 & 0.76 & 0.01 & 0.74 & 0.01 \\
\ + GNN-Coder & \textbf{0.02} & 0.01 & \textbf{0.01} & 0.01 & \textbf{0.02} & \textbf{0.00} & \textbf{0.02} & \textbf{0.00} & \textbf{0.02} & \textbf{0.00} & \textbf{0.02} & \textbf{0.00} \\
\hline

\end{tabular}
\end{adjustbox}

\caption{MAM score comparison for different methods on the CSN dataset, providing insight into the uniformity of code embedding distribution. Lower value indicates better distribution.}
\label{tab:mam}
\end{table*}

\noindent
\textbf{Results on CSN.}
Considering that UniXcoder is pre-trained on the CSN dataset, whereas CodeT5+ incorporates a larger volume of code data, we first assess the effectiveness of GNN-Coder in improving Transformer model performance on the observed datasets.

The experimental results, presented in Table~\ref{tab:csn_mrr_recall}, provide compelling evidence of the effectiveness of the GNN-Coder in the text-to-code retrieval task. Specifically, GNN-Coder consistently enhances the performance of all Transformer-based models, with the most significant improvements observed in models that exhibit lower initial capabilities in code generation. Notably, despite UniXcoder’s partial incorporation of AST information, GNN-Coder still outperforms it. This result can be attributed to the superior ability of GNN-Coder to exploit structured AST information, thereby yielding better performance in the retrieval task.
% further bridging of the gap between pre-training and inference.
% By introducing AST information, GNN-Coder can further enhance the performance of the Transformer model. 

Furthermore, we report the MAM score in Table~\ref{tab:mam}, which provides insight into the uniformity of code embedding distributions. All Transformer models exhibit varying degrees of non-uniformity across datasets. Notably, UniXcoder demonstrates the least non-uniformity, while CodeT5+ shows a higher degree. This discrepancy stems from the fact that CodeT5+ is trained on a broader and more diverse set of code data. while improving performance, it introduces greater complexity and irregularities in embedding distributions. In contrast, the general model LLM-Embedder, with limited capacity to process code-specific features, exhibits a significantly higher level of non-uniformity, highlighting the challenges faced by non-specialized models in handling code data effectively.
The introduction of the GNN model significantly improves embedding uniformity. Despite initial non-uniformity in the output distribution of the Transformer model, the GNN model effectively minimizes these biases, with the average MAM scores of all embeddings approaching zero. This improvement is attributed to the learnable parameter $\lambda$ within the GNN framework, which dynamically adjusts the balance between the Transformer and GNN models. Additionally, the reduction in the variance of most MAM scores indicates more symmetric and reliable embeddings, demonstrating that the GNN model not only enhances uniformity but also improves the stability and consistency of the results.

\begin{table}[t]
\centering
\renewcommand{\arraystretch}{1.0}
\begin{adjustbox}{width=0.5\textwidth}
\begin{tabular}{lcccc}
\hline
Model & MRR & R@1 & Mean & SD \\
\hline
UniXcoder 110M & 27.96 & 17.8 & 0.10 & 0.02 \\
\ + GNN-Coder & \textbf{48.22} & \textbf{36.2} & \textbf{0.01} & 0.02 \\
\hdashline
CodeT5+ 110M & 45.44 & 31.2 & 0.24 & 0.03 \\
\ + GNN-Coder & \textbf{66.50} & \textbf{53.8} & \textbf{0.05} & \textbf{0.01} \\
\hline
LLM-Embedder & 46.10 & 33.4 & 0.77 & 0.01 \\
\ + GNN-Coder & \textbf{67.96} & \textbf{56.4} & \textbf{0.02} & \textbf{0.00} \\
\hline

\end{tabular}
\end{adjustbox}
\caption{Results comparison in terms of MRR, R@1 and MAM on various Transformer architectures on the CosQA dataset.}
\label{tab:cosqa_res}
\end{table}

\noindent
\textbf{Zero-shot Performance on CosQA.}
To evaluate the generalization ability of GNN-Coder, we use CosQA dataset, which is not part of the training set for the code Transformer model, and evaluate the performance improvement of GNN-Coder. The results, presented in Table~\ref{tab:cosqa_res}, show that the general model, LLM-Embedder, outperforms the code model. This suggests that the code model has limited generalization capability, likely due to the significant differences between the CosQA dataset and the training data of the code Transformer model. Notably, after incorporating the GNN model, all models exhibit substantial performance improvements. Furthermore, the reported MAM score indicates that these improvements may be attributed to a more uniform distribution of code.

% This fully demonstrates that GNN-Coder can effectively bridge the domain gap between the training set and the test set by means of the GNN model.

\begin{table*}[t]
\centering
\renewcommand{\arraystretch}{1.0}
\begin{adjustbox}{width=\textwidth}
\begin{tabular}{llccccccc}
\hline
Model & Pooling Layer & Ruby & JavaScript & Go & Python & Java & PHP & Avg \\
\hline
\multirow{3}{*}{UniXcoder 110M} & TopKPool & 49.14 & 39.37 & 67.24 & 42.80 & 46.10 & 35.23 & 46.65 \\
& SAGPool & 50.08 & 39.49 & 66.29 & \textbf{44.90} & 45.83 & 35.16 & 46.96 \\
& ASTGPool & \textbf{50.40} & \textbf{40.22} & \textbf{67.76} & 44.56 & \textbf{46.42} & \textbf{35.86} & \textbf{47.54} \\
\hline
\multirow{3}{*}{CodeT5+ 110M} & TopKPool & 73.72 & 66.93 & 90.25 & \textbf{70.61} & 70.70 & 65.73 & 73.02 \\
& SAGPool & 73.66 & 66.84 & 90.25 & 70.44 & 70.73 & \textbf{66.00} & 72.99 \\
& ASTGPool  & \textbf{73.85} & \textbf{67.20} & \textbf{90.71} & 70.37 & \textbf{70.80} & 65.93 & \textbf{73.14} \\
\hline
\multirow{3}{*}{LLM-Embedder} & TopKPool & 61.62 & 49.31 & 84.65 & 56.84 & 54.33 & 48.95 & 59.28 \\
& SAGPool & 61.35 & 49.40 & 84.63 & 56.70 & 54.86 & 48.75 & 59.28 \\
& ASTGPool  & \textbf{65.00} & \textbf{52.35} & \textbf{87.10} & \textbf{61.71} & \textbf{60.38} & \textbf{54.96} & \textbf{63.58} \\
\hline

\end{tabular}
\end{adjustbox}
\caption{Illustrating the effect of the proposed ASTGPool by comparing different pooling layers in terms of MRR with respect to various Transformer architectures on the CSN dataset. }
\label{tab:pooling_layer}
\end{table*}

\begin{table*}[t]
\centering
\renewcommand{\arraystretch}{1.0}
\begin{adjustbox}{width=\textwidth}
\begin{tabular}{l|cc|cc|cc|cc|cc|cc|c}
\hline
\ & \multicolumn{2}{c|}{Ruby} & \multicolumn{2}{c|}{JavaScript} & \multicolumn{2}{c|}{Go} & \multicolumn{2}{c|}{Python} & \multicolumn{2}{c|}{Java} & \multicolumn{2}{c|}{PHP} & \multicolumn{1}{c}{Avg} \\
 & MRR & Mean & MRR & Mean & MRR & Mean & MRR & Mean & MRR & Mean & MRR & Mean & MRR \\
\hline
CodeT5+ 110M & 73.55 & 0.16 & 65.83 & 0.17 & 89.51 & 0.15 & 69.75 & 0.13 & 69.42 & 0.14 & 64.44 & 0.15 & 72.08\\
\hdashline
+ MLP Adapter & 73.75 & 0.07 & 67.18 & 0.08 & 90.35 & 0.05 & 70.22 & 0.06 & 70.75 & 0.06 & 65.76 & 0.06 & 73.00 \\
+ GNN wo pooling & 73.65 & 0.02 & 67.19 & 0.02 & 90.37 & 0.00 & \textbf{70.50} & 0.01 & \textbf{70.96} & 0.01 & 65.84 & 0.01 & 73.09 \\
+ GNN w pooling (best) & \textbf{73.85} & 0.01 & \textbf{67.20} & 0.02 & \textbf{90.71} & 0.00 & 70.37 & 0.01 & 70.80 & 0.01 & \textbf{65.93} & 0.01 & \textbf{73.14} \\
\hdashline
\ \ \ \ - no AST node type & 73.64 & 0.02 & 66.47 & 0.03 & 89.69 & 0.00 & 69.78 & 0.02 & 70.51 & 0.02 & 65.88 & 0.01 & 72.66 \\
\ \ \ \ - undirect AST & 73.32 & 0.02 & 67.07 & 0.02 & 90.34 & 0.00 & 70.40 & 0.01 & 70.86 & 0.01 & 66.10 & 0.01 & 73.02 \\
\hline
LLM-Embedder & 63.07 & 0.76 & 49.33 & 0.77 & 80.94 & 0.69 & 55.86 & 0.76 & 53.60 & 0.76 & 44.70 & 0.74 & 57.92\\
\hdashline
+ MLP Adapter & 61.53 & 0.19 & 52.05 & 0.19 & 85.87 & 0.27 & 60.41 & 0.26 & 58.16 & 0.22 & 51.72 & 0.18 & 61.62 \\
+ GNN wo pooling & 61.53 & -0.01 & 49.77 & 0.01 & 84.77 & 0.02 & 61.69 & 0.02 & 58.90 & 0.02 & 54.21 & 0.02 & 61.81 \\
+ GNN w pooling (best) & \textbf{65.00} & 0.02 & \textbf{52.35} & 0.01 & \textbf{87.10} & 0.02 & \textbf{61.71} & 0.02 & \textbf{60.38} & 0.02 & \textbf{54.96} & 0.02 & \textbf{63.58} \\
\hdashline
\ \ \ \ - no AST node type & 64.45 & 0.02 & 52.13 & 0.01 & 86.51 & 0.02 & 59.12 & 0.01 & 58.57 & 0.02 & 54.09 & 0.02 & 62.48 \\
\ \ \ \ - undirect AST & 64.78 & 0.01 & 52.25 & 0.01 & 86.88 & 0.02 & 61.51 & 0.02 & 55.00 & 0.01 & 54.80 & 0.03 & 62.54 \\
\hline

\end{tabular}
\end{adjustbox}
\caption{Retrieval results under different settings on the CSN dataset. For experiments removing parts of AST information, we set the pooling ratio to 0.1, which is generally best.}
\label{tab:GNN_ablation}
\end{table*}

\subsection{Ablation Study}

\noindent 
\textbf{The Effect of Pooling Layer.}
We first perform an ablation study on the pooling layers in the GNN model, comparing the performance of TopKPool, SAGPool, and ASTGPool. The results are summarized in Table~\ref{tab:pooling_layer}. As described in Section~\ref{sec:GNN_architecture}, existing pooling methods struggle with reverse AST processing. In contrast, ASTGPool addresses this issue by accounting for both the number of adjacent nodes and their characteristics, thus overcoming the limitations of previous methods. The experimental results demonstrate that ASTGPool outperforms all other pooling methods.

Furthermore, we conduct experiments on GNN-Coder without a pooling layer, as shown in Table~\ref{tab:GNN_ablation}. It indicates that the impact of the pooling layer varies across models. Specifically, the pooling layer has a more pronounced effect on improving the performance of LLM-Embedder compared to CodeT5+. This difference arises from the models' distinct code processing capabilities. For LLM-Embedder, which has relatively limited code processing ability, most AST nodes contain noise, and pooling improves model performance. Conversely, for CodeT5+, which has a stronger code processing ability, most AST nodes carry valuable information, and pooling may result in information loss.

\noindent
\textbf{The Effect of GNN.} 
We conduct an ablation study on GNN-Coder with various configurations. First, we replace the GNN model with a basic Multi-Layer Perceptron (MLP) Adapter, which remaps Transformer embeddings without AST information. Next, we examine the role of AST information by evaluating two variations: excluding AST node types and converting directed edges to undirected edges. The experimental results in terms of MRR and Mean MAM are summarized in Table~\ref{tab:GNN_ablation}.

The study reveals that the MLP Adapter mitigates the non-uniformity of the Transformer model to some extent, leading to performance improvements across all models. However, the enhancement is modest compared to the GNN model, underscoring the critical role of integrating AST information. Additionally, removing the AST information significantly degrades model performance, even below the baseline performance with the MLP Adapter in some cases. This suggests that performance gains are primarily driven by the inclusion of AST data, rather than an increase in model parameters. Overall, the proposed GNN-Coder achieves the best performance by incorporating AST semantic information with a GNN model.

% To evaluate the advantages of uniform feature distribution and incorporation of AST information respectively, we conduct a study on GNN-Coder with different settings. First, we replace the GNN model with a simple Multi-Layer Perceptron (MLP) Adapter. This adapter remaps the Transformer embeddings under the condition of not containing AST information. Next, we investigate the impact of AST information in two ways: one is to exclude the AST node types, and the other is to convert directed edges into undirected edges. The experimental results are presented in Table~\ref{tab:GNN_ablation}.

% The study found that the MLP Adapter alleviates the non-uniformity of the Transformer model to a certain extent, thereby improving the performance of all models. However, compared with the GNN model, this improvement effect is relatively limited. This comparison result highlights the important role of incorporating AST information.
% In addition, when part of the AST information is removed, the model performance drops significantly, even lower than the effect brought by the MLP Adapter. This phenomenon indicates that the improvement of model performance stems from incorporating AST data, rather than merely the factor of increasing model parameters.
% In conclusion, these experimental results demonstrate that GNN-Coder achieves performance improvement through two aspects: incorporating AST semantic information and achieving uniform feature distribution. 

\section{Conclusion}
we propose GNN-Coder, a novel framework for code retrieval that leverages GNNs and ASTs to overcome the limitations of traditional sequence-based models. By integrating GNNs with Transformers, GNN-Coder effectively captures both structural and semantic features of code, addressing the challenges posed by structurally complex code fragments. The introduction of a tailored graph pooling method further enhances the model’s ability to retrieval accuracy through better feature separation. Our experiments demonstrate that GNN-Coder outperforms existing methods, achieving significant improvements in retrieval performance across multiple datasets. The results highlight the potential of GNN-Coder to advance the field of code retrieval, offering a promising solution for handling complex code structures and enhancing the effectiveness of software development tools.

% We propose the novel framework of GNN-Coder. By combining the GNN model with Transformers, this framework effectively improves the performance of Transformer models in the code retrieval task. Specifically, GNN-Coder can significantly reduce the non-uniformity of Transformer models on retrieval datasets. Meanwhile, by leveraging AST information, it further optimizes model performance.
% We conduct experiments on two datasets, CSN and CosQA. The results clearly show that GNN-Coder significantly enhances the performance of all models. This is especially true for models with relatively low code processing capabilities or those facing unseen datasets.
% In the ablation study on the pooling layer, we found that ASTGPool has obvious advantages compared to other pooling methods. It can not only reduce computational complexity but also improve the performance of GNN-Coder when dealing with models with low code processing capabilities. The ablation study on GNN and AST emphasizes the crucial role of AST information in performance improvement. 
% At the same time, it also highlights how bias restricts the effectiveness of Transformer models. 

\section{Limitation}
The GNN-Coder framework presented in this work is primarily designed for the code retrieval task, demonstrating its potential within this domain. However, the framework's applicability could extend to other related tasks, such as code clone detection and code translation. Future research should explore a broader range of tasks to better assess the framework's effectiveness across various code-related applications.

A limitation in the current approach lies in the handling of text embeddings. As detailed in Appendix~\ref{app:SD'}, the uniform distribution of code embeddings does not entirely eliminate slight non-uniformity in text embeddings, caused by the dynamic nature of the distribution metrics. Given that GNN-Coder is primarily tailored for code modality, further investigation into the text embedding non-uniformity is necessary. Future work will focus on addressing this issue and evaluating the framework's generalization capabilities, particularly in combination with LLMs.

Additionally, the experiments conducted so far have been limited to models with a relatively small parameter scale due to hardware constraints. To fully understand the scalability of the GNN-Coder framework, future work should explore its performance and applicability on larger models and more diverse types of model architectures.

% The GNN-Coder framework we proposed is mainly oriented towards code retrieval task. It is worth noting that this framework may also have broad application prospects in other similar tasks such as code clone detection and code translation. In future research, experiments on more types of tasks can be considered to more effectively verify the effectiveness of the GNN-Coder framework in various code-related tasks.
% In addition, as discussed in Appendix~\ref{app:SD'}, after code embeddings achieve a uniform distribution, text embeddings may exhibit slight non-uniformity due to the dynamic and mutual nature of the distribution metrics. As GNN-Coder is designed primarily for code modality, further investigation into the non-uniformity in text embeddings is needed. Future work will focus on exploring this non-uniformity and evaluating GNN-Coder’s generalization ability with LLM models. Finally, due to the limitations of hardware conditions, we have conducted experimental verification on models with a relatively small parameter scale only. Future work can explore the applicability and performance of this framework in more types of models and larger models.

\bibliography{custom}

\appendix

\section{Training Process}
\label{app:training_process}

During the training process, we choose the AdamW optimizer and set the learning rate to 0.004. We warm up the learning rate in the first 10\% of the training steps and then adopt a cosine annealing decay strategy. Specifically, for Transformer models with an output dimension of 256, we train the GNN model for 400 iterations with a batch size of 16K. For models with an output dimension of 768, to ensure the training effect, we adjust the batch size to 8K and increase the number of training iterations to 1000.

Notably, for UniXcoder and CodeT5+, different from the original settings, we preserve the original format of the code before tokenization instead of splitting each word. This is because preserving the original code format is more reasonable in real-world application scenarios. Therefore, the experimental results of UniXcoder and CodeT5+ may slightly differ from the original reports.

\section{Pooling Ratio} We also investigate the performance of different pooling ratios, varying the ratio from 0.1 to 0.9. The results on CSN and CosQA datasets are presented in Table \ref{tab:pooling_ratio}. For CodeT5+, which exhibits the best inherent code capability, the performance is relatively insensitive to the pooling ratio, with optimal results observed at ratios of 0.1 and 0.7. This insensitivity may be due to the well-initialized AST, where most nodes are significant. As a result, while reducing computational complexity, the pooling layer may slightly impact performance due to information loss, as detailed in Table \ref{tab:GNN_ablation}. For the other two models, which have lower code capabilities, the initialized AST has more noise. Therefore, a lower pooling ratio is more effective in reducing noise and enhancing performance, with the optimal ratio being 0.1. In these cases, the pooling layer plays a more significant role in improving performance.

\begin{table*}[t]
\centering
\renewcommand{\arraystretch}{1.0}
\begin{adjustbox}{width=\textwidth}
\begin{tabular}{l|c|ccccccc|c}
\hline
model & Pooling Ratio & \multicolumn{7}{c|}{CSN} & CosQA \\
\ & \ & Ruby & JavaScript & Go & Python & Java & PHP & CSN\_Avg  \\
\hline
\multirow{5}{*}{UniXcoder 110M} & 0.1 & \textbf{50.40} & \textbf{40.22} & \textbf{67.76} & \textbf{44.56} & \textbf{46.42} & \textbf{35.86} & \textbf{47.54} & \textbf{48.22} \\
& 0.3 & 50.18 & 39.54 & 67.31 & 44.28 & 46.00 & 35.30 & 47.21 & 48.20 \\
& 0.5 & 49.75 & 39.78 & 67.07 & 43.93 & 46.10 & 35.25 & 46.98 & 47.16 \\
& 0.7 & 49.69 & 39.57 & 66.65 & 44.07 & 46.11 & 35.50 & 46.93 & 47.22 \\
& 0.9 & 49.73 & 39.68 & 65.63 & 43.57 & 45.97 & 35.33 & 46.65 & 46.64 \\
\hline
\multirow{5}{*}{CodeT5+ 110M} & 0.1 & 73.69 & 67.15 & \textbf{90.71} & \textbf{70.37} & 70.77 & 65.91 & \textbf{73.10} & 65.51 \\
& 0.3 & 73.72 & 67.08 & 90.48 & 70.36 & 70.76 & 65.93 & 73.06 & 64.51 \\
& 0.5 & 73.64 & 67.08 & 90.46 & 70.37 & 70.78 & 65.90 & 73.04 & 64.39 \\
& 0.7 & \textbf{73.85} & \textbf{67.20} & 90.47 & 70.36 & \textbf{70.80} & \textbf{65.93} & \textbf{73.10} & \textbf{66.50} \\
& 0.9 & 73.70 & 67.10 & 90.51 & 70.36 & 70.76 & 65.91 & 73.06 & 65.36 \\
\hline
\multirow{5}{*}{LLM-Embedder} & 0.1 & \textbf{65.00} & \textbf{52.35} & \textbf{87.10} & \textbf{61.71} & \textbf{60.38} & \textbf{54.96} & \textbf{63.58} & \textbf{67.96} \\
& 0.3 & 62.80 & 51.05 & 86.92 & 61.30 & 60.32 & 54.94 & 62.89 & 67.50 \\
& 0.5 & 62.63 & 50.66 & 86.92 & 61.59 & 60.32 & 54.76 & 62.81 & 66.50 \\
& 0.7 & 62.18 & 50.54 & 87.07 & 61.63 & 60.27 & 54.66 & 62.73 & 67.41 \\
& 0.9 & 61.73 & 50.35 & 86.97 & 61.40 & 60.35 & 54.45 & 62.59 & 67.60 \\
\hline

\end{tabular}
\end{adjustbox}
\caption{MRR of different pooling ratios with various models and datasets.}
\label{tab:pooling_ratio}
\end{table*}

\section{Distribution of Text Embeddings}
\label{app:SD'}

To evaluate the distribution of the text embeddings, we also calculate the standard deviation (SD) of $\text{MAM}^{\prime}$:
\begin{equation}
\text{MAM}^{\prime}_i = \frac{1}{N} \sum_{j=1}^{N} \text{sim}(\mathbf{c}_i, \mathbf{t}_j)
\end{equation}

As shown in Table~\ref{tab:SD'_values}, we observe that after incorporating the GNN model, the SD of some MAM$^{\prime}$ remains the same or even increases. The decrease in the SD of MAM is due to all MAM approaching zero. The increase in the SD of MAM$^{\prime}$ is more complex. After adding the GNN model, the code embeddings become more dispersed, leading to more diverse observation angles for the text encoded space. However, since the text embeddings are not modified, their SD may slightly increase. Finally, the SD of MAM$^{\prime}$ are even higher than those of MAM, suggesting that the distribution of text embeddings is less uniform compared with code embeddings. This may be because text embeddings share the encoding space with all texts, which may introduce a certain degree of non-uniformity. Given that GNN-Coder is mainly designed for the code modality, future research should focus on the non-uniformity issues in text embeddings.

\begin{table*}[t]
\centering
\renewcommand{\arraystretch}{1.0}
\begin{adjustbox}{width=0.9\textwidth}
\begin{tabular}{l|cccccc|c}
\hline
Model & \multicolumn{6}{c|}{CSN} & CosQA \\
\ & Ruby & JavaScript & Go & Python & Java & PHP \\
\hline
UniXcoder 110M & 0.03 & 0.03 & 0.02 & 0.03 & 0.03 & 0.04 & 0.03 \\
\ + GNN-Coder & \textbf{0.02} & \textbf{0.01} & 0.02 & \textbf{0.02} & \textbf{0.02} & \textbf{0.02} & \textbf{0.02} \\
\hdashline
CodeT5+ 110M & 0.02 & 0.02 & 0.02 & 0.02 & 0.02 & 0.02 & 0.05 \\
\ + GNN-Coder & 0.02 & 0.02 & 0.02 & 0.02 & 0.02 & 0.02 & \textbf{0.02} \\
\hline
LLM-Embedder & 0.01 & \textbf{0.01} & \textbf{0.01} & \textbf{0.01} & 0.01 & \textbf{0.01} & 0.02 \\
\ + GNN-Coder & 0.01 & 0.02 & 0.02 & 0.02 & 0.01 & 0.02 & 0.02 \\
\hline

\end{tabular}
\end{adjustbox}
\caption{The distribution of the text embeddings by different models on different datasets. The data in the table represents the standard deviation of MAM$^{\prime}$. The mean value of MAM$^{\prime}$ is the same as that of MAM.}
\label{tab:SD'_values}
\end{table*}

\end{document}